\documentclass[aps,prl,reprint]{revtex4-1}
\usepackage{graphicx}

\begin{document}
\title{Ordering Dynamics in Neuron Activity Pattern Model: An insight to Brain
Functionality}
\author{Awaneesh Singh$^{1}$, Jasleen Gundh$^2$, and R.K. Brojen Singh$^2$}
\email{brojen@jnu.ac.in}
\affiliation{$^1$School of Physical Sciences, Jawaharlal Nehru University, New Delhi$-$110067, India.\\
$^2$School of Computational and Integrative Sciences, Jawaharlal Nehru University, New Delhi-110067, India.}

\begin{abstract}
We study the ordering kinetics in $d=2$ ferromagnets which corresponds to populated neuron activities with long-ranged interactions, $V(r)\sim r^{-n}$ associated with short-ranged interaction. We present the results from comprehensive Monte Carlo (MC) simulations for the nonconserved Ising model with $n\ge 2$. Our results of long-ranged neuron kinetics are consistent with the same dynamical behavior of short-ranged case ($n > 4$). The calculated characteristic length scale in long-ranged interaction is found to be $n$ dependent ($L(t)\sim t^{1/(n-2)}$), whereas short-ranged interaction follows $L(t)\sim t^{1/2}$ law and approximately preserve universality in domain kinetics. Further, we did the comparative study of phase ordering near the critical temperature which follows different behaviours of domain ordering near and far critical temperature but follows universal scaling law.
\end{abstract}


\maketitle

\section{Introduction}
\label{Intro}
The spiking activity in complex neuron network in brain is dynamic (far from equilibrium) \cite{liu}, exhibit nonequilibrium critical dynamics \cite{lo} and the criticality in it has been used to characterize brain signals \cite{chia}. The reason could be when such system is quenched from a homogeneous phase to a broken-symmetry phase, it becomes thermodynamically unstable. The subsequent {\it far-from-equilibrium} evolution of the system is characterized by the emergence and growth of domains enriched in the new equilibrium phases. This nonequilibrium evolution, usually called {\it kinetics of phase ordering} or {\it domain growth}, has been the subject of much active investigation \cite{pw09}. The domain morphology is quantified by the time dependence of the domain scale $L(t)$, where $t$ is the time after the quench. There is a good understanding of domain growth kinetics in pure and isotropic systems with short-ranged interactions, where the domain scale shows a power-law behavior, $L(t) \sim t^{\phi}$ \cite{ab94, ao02}. For the case with nonconserved order parameter, e.g., ordering of a ferromagnet into up and down phases (spin-flip Glauber-Ising model \cite{rg63}), one has $\phi = 1/2$ \cite{il62, ac79}. On the other hand, for the case with conserved order parameter, e.g., phase separation of a binary $(AB)$ mixture into $A$- and $B$-rich domains (spin-exchange Kawasaki-Ising model \cite{kk66}), we have $\phi = 1/3$ when growth is driven by diffusion \cite{ab89, ab90}. Apart from the domain growth laws, experimentalists are also interested in quantitative features of the domain morphologies. An important experimental quantity is the time-dependent correlation function $C(\vec{r},t)$ or its Fourier transform, structure factor $S(\vec{k},t)$ ($\vec{k}$ being the wave vector \cite{pw09, ab94}. Most of the studies are concentrated around the nearest-neighbor (short-range) inter-molecular interaction. 

The neuron activities in brain at critical point are believed to be effective for the long distance communication of the neurons \cite{begg} because of coupling and variability to optimize information storage in the system \cite{soco} and dynamic range of the system to response the input signal \cite{kino}. The core of the paper focuses on long-range interactions which dynamically explains the rapid movement of the signal information inside the brain. The fast emergence of long-range interactions, mimic the rapid neuronal interactions in the brain. Further, we study the phase ordering dynamics in neurons with a specific interest to understand the role of the range of inter-neuron interactions. We address two important questions in this context via kinetic MC simulations:
(a) What is the growth law for ordering phases of neurons? Is the growth law independent of the range of interaction?
(b) What is the morphology for ordering phases of neurons, as measured by the correlation function and structure factor? Is it comparable for all interaction range? We will be providing the answers to the above questions from our extensive MC simulations.

\section{Neuron Activity Pattern Model}
\label{mds}
Brain can be considered as a complex network of neurons which can be mapped onto ferromagnetic Ising model with two spin interactions \cite{tou}, where random firing or non-firing of neuron can be represented by two states of a spin, $s=+1$ for firing and $s=-1$ for rest or non-firing neurons \cite{fra}. Even though neuron activity pattern model is far from equilibrium dynamic model \cite{schn}, Ising model can be serve an excellent model to deal with critical phenomena of neuron activity pattern \cite{liu}. The large number of local (short range) interaction of neurons \cite{bass} and significant amount of global (long range) interaction of neurons \cite{bre} are main basis of neurons communication in brain network \cite{sch}. We consider the following Hamiltonian of two dimensional Ising system which incorporates long-ranged spin (neuron) model (LSM),
\begin{eqnarray}
 H = -\displaystyle\sum_{<ij>}J(r_{ij},n) s_is_j, \quad s_i=\pm 1,
\label{IH}
\end{eqnarray}
where $J$ is the coupling strength, $n$ characterizes the range of the interaction, $r_{ij}=|\vec{r}_i - \vec{r}_j|$, and $s_i$ denotes the spin variable at site $i$. We consider two state spins: $s_i = +1$ denotes an up-spin (active neurons) and $s_i = -1$ denotes a down-spin (passive neurons). We consider only a ferromagnetic case, where $J > 0$ always. The case where $J$ can be both $> 0$ (ferromagnetic) and $< 0$ (antiferromagnetic) is relevant to spin glasses. We associate stochastic dynamics with the Ising model by placing it in contact with a heat bath. The appropriate dynamics for the phase ordering problem is {\it spin-flip kinetics} or {\it Glauber kinetics}.
\begin{figure}[htbp]
\centering
\includegraphics[width=3.2in]{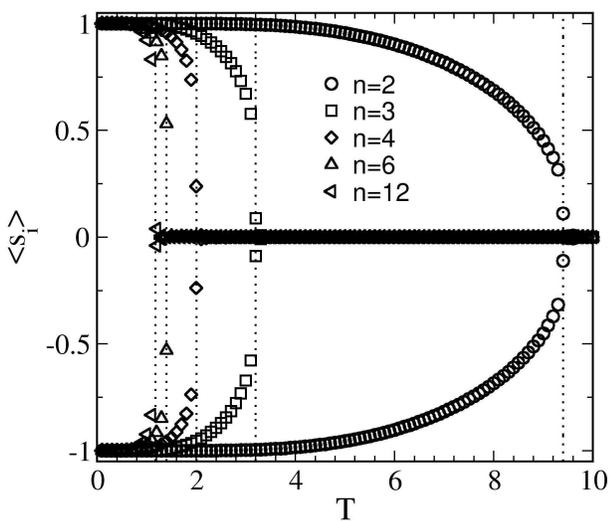}
\caption{Plot of $<s_i>$ vs. $T$ for $n$= 2, 3, 4, 6, and 12 as indicated. The magnetization drops-off sharply near the critical temperature ($T_c$) and then vanishes to $0$ in the disordered high-temperature phase.}
\label{fig1}
\end{figure}

If we consider interacting neurons through slowly decay potentials \cite{lamo}. In order to capture thermodynamical parameters, one can define the following potential function (obeying power law functional form) \cite{hil,cann}, 
\begin{eqnarray}
\label{un}
U(n)=\lim_{N\rightarrow\infty}\frac{1}{N}\sum_{i,j;i\ne j}^N\frac{J}{r_{ij}^{n}}.
\end{eqnarray}
Here $n=d+\sigma=2+\sigma$ \cite{fish} for two dimensional system. For short-ranged interaction, $\sigma>2$ and the system size is not much important, whereas for long-ranged interaction, $0<\sigma<2$ and it depends on the system size. Since, the size of the system is $N$, rescaling $J\rightarrow J/N$ to the Curie-Weiss model \cite{cann} and using Euler-McLaurin sum formula \cite{brui} for $N\gg 1$, equation (\ref{un}) can be written as,
\begin{eqnarray}
\label{unn}
U(n)&=&\lim_{N\rightarrow\infty}J\sum_{x=1}^{\sqrt{N}}\sum_{y=1}^{\sqrt{N}}\frac{1}{\left(x^2+y^2\right)^{(n-2)/2}},\nonumber\\
&\approx& \lim_{N\rightarrow\infty}J2^{d(=2)}\int_1^{\sqrt{N}}drg(r)r^{3-n},
\end{eqnarray}
where, $g(r)$ is the pair distribution function such that $g(r)\approx 1$ for $r\gg 1$. Then the equation (\ref{unn}) becomes,
\begin{eqnarray}
U(n)\approx \lim_{N\rightarrow\infty}J
\left\{\begin{array}{lll}2\ln(N)&\mbox{for}&n=4,\\
\frac{4}{n-4}\left(1-N^{2-n/2}\right)&\mbox{for}&n > 4,\\
\frac{1}{1-n/4}N^{2-n/2}&\mbox{for}&0\le n < 4
\end{array}\right.
\label{qn}
\end{eqnarray}
From equation (\ref{qn}), one can see that $U(n)$ is finite for $n\ge 4$ when $N\rightarrow\infty$, and the asymptotic behaviour of finite critical temperature $T_c$ \cite{hil},
\begin{eqnarray}
\label{tc1}
T_c(n)\approx \frac{J}{k_B}U(n)\approx \frac{J}{k_B} \left(\frac{4}{n-4}\right),
\end{eqnarray}
where, $k_B$ is Boltzmann constant. This shows that $T_c(n)\propto 1/n$ for short-ranged potential ($n>4$), whereas for long-ranged potential, $T_c$ depends on the size of the system $N$ as well as $n$ given by,
\begin{eqnarray}
\label{tc2}
T_c(n,N)\approx \frac{J}{k_B}\left(\frac{4}{4-n}\right)N^{2-n/2},
\end{eqnarray}
and $T_c$ diverges with system size. Similarly, one can also calculate other thermodynamical parameters such as internal energy, entropy, free energy per particle (neuron) etc. at this asymptotic limit.
\begin{figure}[htbp]
\centering
\includegraphics[width=3.2in]{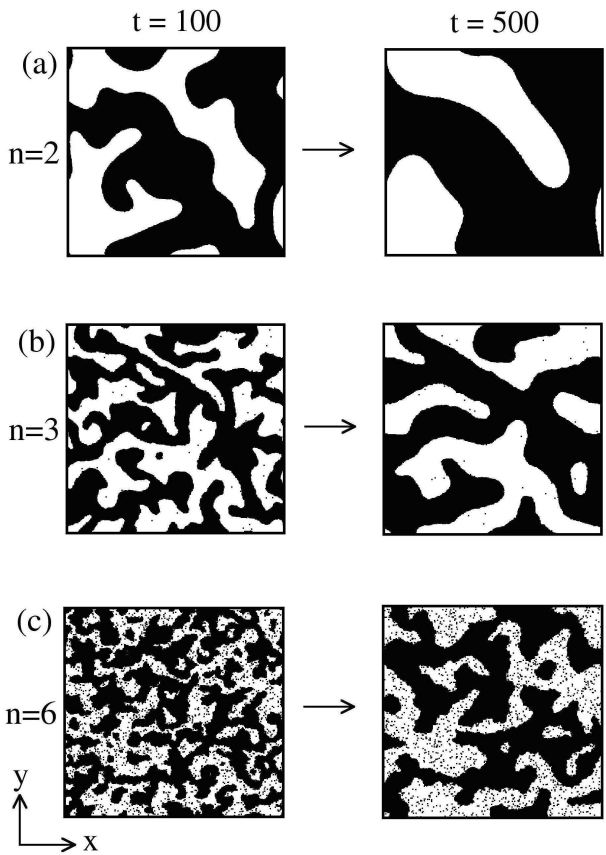}
\caption{Evolution snapshots of domain coarsening for $n=2$, 3, and 6 quenched at $T=1$ below $T_c$ at time $t=100$, 500. The snapshots are obtained from a Monte Carlo (MC) simulation of ordering kinetics in ferromagnetic system. The details of the MC simulation are provided in the text.}
\label{fig2}
\end{figure}

\section{Details of Simulation}

Since it is very difficult to obtain exact analytical solution of this problem, we straightforward implement a MC simulation of the Ising model with spin-flip kinetics to understand the behaviour. In a single step of MC dynamics, we choose a spin at random in the lattice of distribution of spins. The change in energy $\Delta H$ that would occur if the spin was flipped is computed with the step of acceptance or rejection based on Metropolis acceptance probability \cite{bh02,nb99} given by,
\begin{eqnarray}
P=\left\{\begin{array}{ll} \exp(-\beta \Delta H) & \mbox{if} ~\Delta H \ge 0,\\ 1 & \mbox{if} ~\Delta H < 0. \end{array} \right.
\label{P}
\end{eqnarray}
where, $\beta={(k_{B}T)}^{-1}$ denotes the inverse temperature. One Monte Carlo step (MCS) is completed when this algorithm is performed $N$ times (where $N$ is the total number of spins), regardless of whether the move is accepted or rejected. All our simulations have been performed on a $d=2$ lattice of size $L_s^2$ ($L_s=512$) with periodic boundary conditions in both directions. The statistical quantities presented here (e.g., correlation function, structure factor) are obtained as averages over $10$ independent runs. Each run starts with a randomly-mixed state with equal numbers of up (fired) and down (inactive) spins (neurons), which corresponds to a mean magnetization $m=\langle s_i\rangle = 0$. 
\begin{figure}[htbp]
\centering
\includegraphics[width=3.2in]{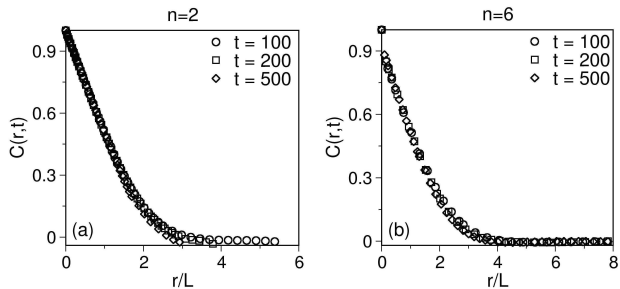}
\caption{(a) Scaling plot of $C(r,t)$ vs. $r/L$ for a phase ordering dynamics in $d=2$ for $n=2$. The data sets (for $t=100,200,500$) collapse onto a single master curve. (b) Similar plot for $n=6$.}
\label{fig3}
\end{figure}

We study LSM for several values of $n$, namely 2, 3, 4, 6, and 12. The critical ordering temperatures, $T_c(n)$ for each $n$ case is shown in Fig.~\ref{fig1}, where the characteristic behavior of spontaneous magnetization ($<s_i>$) is plotted against temperature ($T$). As expected, $T_c(n)$ (dotted lines) increases with decreasing $n$ as evident from equation (\ref{tc1}). Above $T_c$ the spontaneous magnetization vanishes, whereas below $T_c$ it takes a nonzero value, inducing the typical behavior of a ferromagnet. Therefore, the physical properties of such systems and so its phase states depend on the value for the magnetization, the parameter which is termed as {\it order parameter}: an ordered phase in which the spins are aligned appears when $m \neq 0$, while $m = 0$ implies a disordered (or symmetric) phase. Since, $T_c$'s for $n\ge 4$ are very close to each-other and hence, exhibit qualitatively similar behavior (explained shortly). We thus consider $n<4$ cases for the long-ranged interaction. For each value of $n$, we cut-off the interaction at $r_c = (2.5)^{6/n}$ to accelerate our simulation \cite{sp13}. We stress that the simulations are numerically very demanding for larger cut-offs. We compute several statistical quantities to characterize the system. These are described as follows. 
\begin{figure}[htbp]
\centering
\includegraphics[width=3.0in]{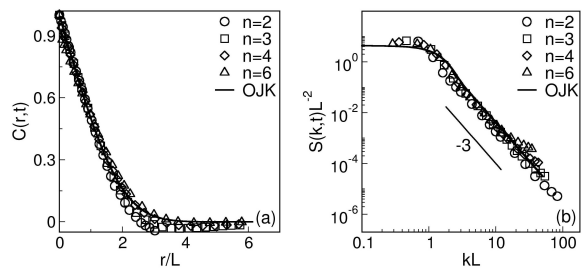}
\caption{(a) Plot of $C(r,t)$ vs. $r/L$ at $t=500$ for $n=2,3,4,6$. (b) Plot of $S(k,t)L^{-2}$ vs. $kL$, corresponding to the data sets in (a). The reasonably good data collapse shows that the scaling functions do not depend on the interaction range. The solid line denotes the OJK function in Eq.~(\ref{ojk}) \cite{ojk82}.}
\label{fig4}
\end{figure}

The domain coarsening is characterized by a growing time-dependent length scale $L(t)$. The domain morphology does not change with time, apart from a scale factor. A direct consequence of the existence of a unique length scale is that the system exhibits a dynamical-scaling in the correlation function and structure factor. We compute the time-dependent correlation function:
\begin{eqnarray}
C\left(\vec{r}_i,\vec{r}_j;t\right) \equiv \left\langle {s_i}{s_j}\right\rangle - \left\langle {s_i}\right\rangle \left\langle {s_j}\right\rangle.
\label{C}
\end{eqnarray}
Here, the angular brackets denote an averaging over the independent initial ensemble and different noise realizations. As the system is translationally invariant, the correlation function depends only on $\vec{r}=\vec{r}_j - \vec{r}_i$: 
\begin{eqnarray}
C\left(\vec{r}_i,\vec{r}_j;t\right)= C\left(\vec{r}_i,\vec{r}_i +\vec{r};t\right)=C\left(\vec{r},t\right).
\end{eqnarray}
Usually most experiments study the structure factor, which is the Fourier transform of the real-space correlation function:
\begin{eqnarray}
S\left(\vec{k},t\right) = \displaystyle\int d\vec{r}\;C\left(\vec{r},t\right)e^{i\vec{k}\cdot\vec{r}}.
\label{S}
\end{eqnarray}
Since the system is isotropic, we can improve statistics by spherically averaging the correlation function and the structure factor. The corresponding quantities are denoted as $C\left(r,t\right)$ and $S\left(k,t\right)$, respectively. The correlation function and structure factor obey the dynamical scaling forms:
\begin{eqnarray}
C(r,t) &= g[r/L(t)], \nonumber \\  
S(k,t) &= L(t)^d f[kL(t)].
\label{scale} 
\end{eqnarray}
Here, $g(x)$ and $f(p)$ are scaling functions; $r$ is the separation between two spatial points; $k$ is the magnitude of the wave vector; and $d$ is the system dimensionality.
The characteristic domain size $L(t)$ is obtained as the distance over which the correlation function decays to some fraction (say half) of its maximum value [$C(r,t)=1$ at $r=0$]. There are several other suitable definitions for computing $L(t)$, e.g., first zero-crossing of $C(r,t)$, inverse of the first moment of $S(k,t)$. In the scaling regime, all these definitions differ only by constant multiplicative factors \cite{op87, op88, po88}.  
\begin{figure}[htbp]
\centering
\includegraphics[width=3.2in]{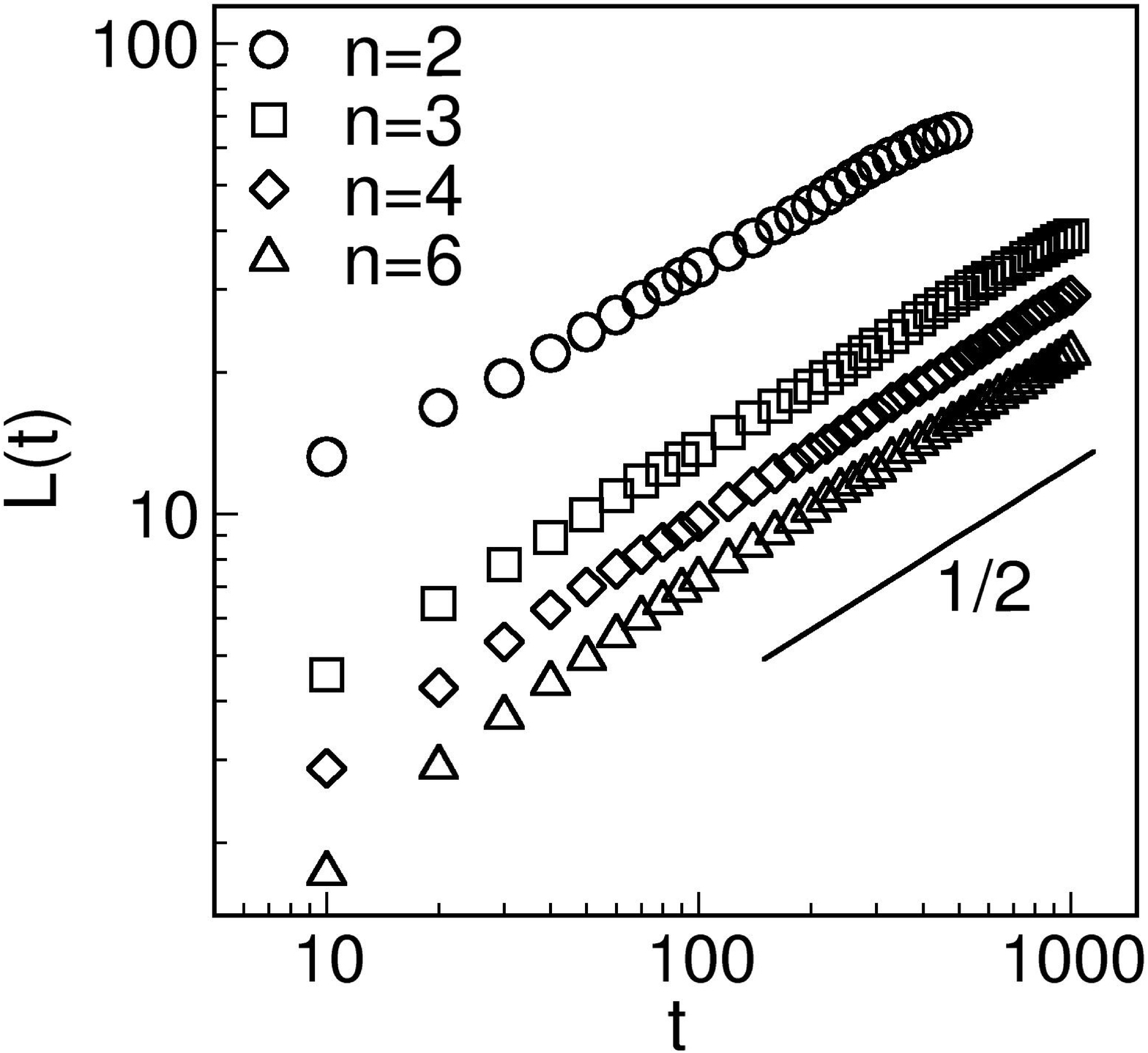}
\caption{Time-dependence of the characteristic length scale $L(t)$ for $n=2,3,4,6$, plotted on a log-log scale. The lines of slope 1/2 indicates the power-law growth regimes expected for phase ordering in $d=2$ ferromagnetic system.}
\label{fig5}
\end{figure}

\section{Numerical Results}
\label{results}
In Fig.~\ref{fig2}, we show the evolution snapshots obtained from our MC simulations for $n=2$, 3, 6 with $T=1$ ($<T_c$, see Fig.~\ref{fig1}) at $t=100$, 500 MCS. At low temperatures, energetic effects are dominant and the system minimizes its energy by ordering the spins parallel to each other. In the absence of an external field (e.g., magnetic field, $h=0$), the activated neuron (up-spin) and inactivated neuron (down-spin) states are equivalent. In the mean-field (MF) limit, i.e., $n = 0$, all the spins interact with each other and there is no spatial structure in the evolution morphology. For larger values of $n$, we see the emergence and growth of domains of up-spin (marked in black) and down-spin (unmarked). These domains interact and annihilate, resulting in coarsening of the characteristic length scale, and therefore, domain patterns at different times look statistically similar, apart from a global change of scale. The domain size at a fixed time (e.g., $t=500$) is smaller for larger values of $n$.

In Fig.~\ref{fig3}, we show a scaling plot of the correlation function, defined in Eq.~(\ref{C}). We plot $C(r,t)$ as a function of the scaled distance $r/L$ at three times, as indicated. Figure~\ref{fig3}(a) corresponds to $n=2$, and Fig.~\ref{fig3}(b) shows data for $n=6$. The data sets show a neat scaling collapse, confirming the validity of dynamical scaling.
\begin{figure}[htbp]
\centering
\includegraphics[width=3.2in]{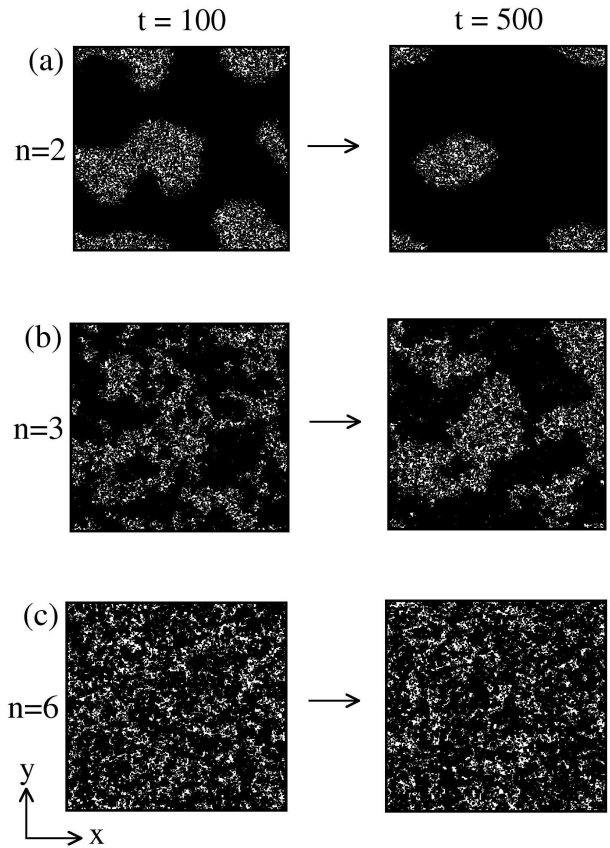}
\caption{Evolution snapshots of phase ordering systems in $d=2$ for $n=2$, 3, and 6. The system is quenched at $T\simeq T_c$. The MC simulations are described in the text.}
\label{fig6}
\end{figure}

Let us next discuss whether the evolution morphology depends on the range of the interaction characterized by $n$. Fig.~\ref{fig4} shows a comparison of the scaling functions for four different $n$ values ($n = 2$, 3, 4, 6) at a time $t=500$, when the system is already in the scaling regime. In Fig.~\ref{fig4}(a), we plot the scaled correlation functions. The reasonably good data collapse suggests that the scaling functions do not depend on the interaction range. The solid line in Fig.~\ref{fig4}(a) denotes the analytical result due to Ohta et al. (OJK) \cite{ojk82}, who studied ordering dynamics in a ferromagnet. The OJK function is
\begin{eqnarray}
C(r,t) = \frac{2}{\pi}\sin^{-1}(e^{-r^2/L^2}).
\label{ojk}
\end{eqnarray}
(The corresponding result for the case with vector order parameter has been obtained by Bray and Puri \cite{bp91}.) Our correlation-function data is in excellent agreement with the OJK function, showing that the phase ordering dynamics for $n<4$ lie in the same dynamical universality class as that for $n > 4$. In Fig.~\ref{fig4}(b), we plot the scaled structure factor [$L^{-2}S(k,t)$ vs. $kL$] for the same time as in Fig.~\ref{fig4}(a). Again, the data sets collapse neatly onto a single master curve, confirming the scaling form in Eq.~(\ref{scale}). The scaling function is in excellent agreement with the corresponding OJK function. Notice that the structure factor, for large values of $k$, follows the well-known Porod's law, $S(k,t)\sim k^{-(d+1)}$, which results from scattering off sharp interfaces \cite{p88, op88a}. The scaled correlation function and structure factor, in congruence with scale free behaviour of functional brain networks \cite{egu} depicts the universality of the interaction mechanism in both short and long range interactions in brain.

In Fig.~\ref{fig5}, we turn our attention to the time-dependence of the domain size. We plot $L(t)$ vs. $t$ on a log-log scale for $n=2$, 3, 4 and 6. Here, the data sets are consistent with the {\it Cahn-Allen} growth law, $L(t)\sim t^{1/2}-$there is no sign of a crossover in the growth law at $n = 4$, as predicted by Bray \cite{ab93}. Bray has used the renormalization group (RG) approach to study ordering dynamics with long-ranged interactions of the form $r^{d+\sigma}$ with $0<\sigma<2$. In our case, $d=2$ and $\sigma=n-2$. Bray argues that the long-ranged interactions are relevant for $0<\sigma<2$ or $2<n<4$, and irrelevant for $n>4$. The corresponding growth law is
\begin{eqnarray}
 L(t) \sim \left\{\begin{array}{l l}
      t^{1/(n-2)}\quad &\text{for}\quad 2<n<4, \nonumber \\
      t^{1/2}\quad &\text{for}\quad n>4, 
\end{array}\right.
\label{bl}
\end{eqnarray}
with possible logarithmic corrections. As we can see that our numerical results are not consistent with this prediction. The only difference as $n$ is varied is that we have faster growth (higher prefactors) for smaller $n$, corresponding to more long-ranged interactions. The fast dynamics of long-range interactions signifies the path of information processing and neuronal connections in the brain. The longer persistance of long-ranged neural connections could give sense to clustering behaviour of neural circuitry , specifically during learning of a specific task, new synaptic connections tend to form in vicinity of old connections related to that task \cite{mfu} making it more robust.

In Fig.~\ref{fig6}, we show the evolution of the order parameter ($m$) near critical temperature ($T\simeq T_c$) from a disordered initial state ($m=0$) for $n=2$, 3, and 6 respectively. At higher temperature below $T_c$ we observe large fluctuations in the evolution patterns with very small global ordering; instead of picking one of the up-spin, down-spin, or zero order parameter states, the system near $T_c$ is a kind of fractal blend of all three \cite{mp98}. However the cluster size is larger for smaller $n$. Recall that at $T=1$ $(\ll T_c)$, thermal energy ($k_BT$) of the system is low, thus spins try to obtain minimal energy by forming domains with a global ordering: $m = +1$ or -1. 
\begin{figure}[htbp]
\centering
\includegraphics[width=3.2in]{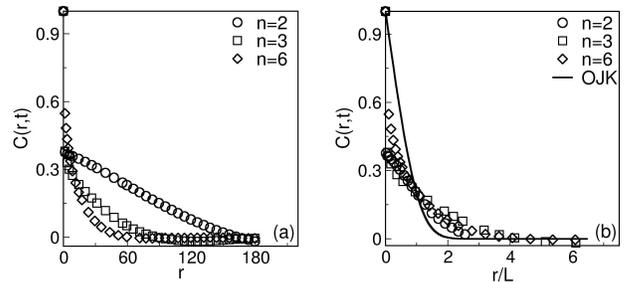}
\caption{(a) Plot of $C(r, t)$ vs. $r$ for $n = 2$, 3, 6 at $t = 500$ MCS. (b) Sacling plot of $C(r,t)$ vs. $r/L$, corresponding to the data sets in (a). The solid line denotes the OJK function for the scaled correlation function in Eq.~(\ref{scale}).}
\label{fig7}
\end{figure}

Finally, in Fig.~\ref{fig7}(a), we show the plot of correlation function [$C(r,t)$ vs. $r$] corresponding to the evolution shown in Fig.~\ref{fig7} at $t=500$. Note that the decay range of the correlation function is larger for smaller $n$. Figure~\ref{fig7}(b) shows the scaling plot of the data sets in (a). A reasonable data collapse confirms the dynamical scaling and clarifies that the system for each $n$ belongs to the same dynamical universality class. 

\section{Summary and Discussion}
\label{summary}
Let us conclude this paper with a summary and discussion of the results presented here. We study the effect of interaction range on the morphology of neuron activity pattern. The long-ranged and short-ranged interaction of neurons could be the main basis of how brain performs complicated functions at fundamental level. Neuron activities as well as wiring and rewiring of the neurons in the network subjected to heat bath depend on the range of interaction which are reflected in the dependence of critical temperature $T_c$ and magnetization on $n$. The domain sizes of the neurons in short range interaction at far-lower critical temperature are smaller; some are isolated and numbers are more as compared to long ranged interaction for any time domain. However, the domain dynamics both in short and long ranged interaction system is quite different as compared to far critical temperature dynamics due to emergence of more randomness in the domain organization in the system. This leads to the change in domain growth laws of the neurons in short and long ranged interaction in the system. The correlation of neurons decays much faster in short ranged interaction as compared to long ranged, but it scales with $r/L$ showing the universality of neuron interaction in brain.

Given the current focus on the biological network and their functionality, we hope that this paper will motivate fresh interest in the evolution dynamics and morphology of active and passive neurons. These kinetic processes play an important role in determining the functionality of brain. We emphasize that one can gain a good understanding of the relevant neuron dynamics (wiring and rewiring inside the brain network) from simple coarse-grained models of the type discussed here. Convincing to the fact that re-learning help us to memorize things for longer duration. Thus it is an attempt to predict and an outlook to understand the functionaliies of the brain.

\vspace{0.5cm}
\noindent {\bf Acknowledgments} \\
JG acknowledges the fellowship and RKBS for providing financial support from the CSIR, India, under sanction no. 25(0221)/13/EMR-II.

\newpage

\end{document}